\documentstyle[12pt,titlepage,epsf]{article}  
\newcommand{\nin}{\noindent}
\newcommand{\be}{\begin{equation}}
\newcommand{\ee}{\end{equation}}
\newcommand{\bea}{\begin{eqnarray}}
\newcommand{\eea}{\end{eqnarray}}

\newcommand{\nonu}{\nonumber\\}

\newcommand{\ol}{\overline}

\begin{document}
\thispagestyle{empty}

\hfill {NTUA-101/00}

\vspace{2cm}

\begin{center}
{\bf\Large{Magnetic catalysis in $QED_3$ 
at finite temperature: \\
beyond the constant mass approximation}}

\vspace{3cm}

{\bf J.Alexandre{\footnote {jalex@central.ntua.gr}}, 
K. Farakos{\footnote {kfarakos@central.ntua.gr}}, 
G. Koutsoumbas{\footnote {kutsubas@central.ntua.gr}}

Department of Physics, National Technical University of
Athens,

Zografou Campus, 157 80 Athens, GREECE}

\vspace{5cm}

Abstract
\end{center}

We solve the Schwinger-Dyson equations for (2+1)-dimensional QED in the
presence of a strong external magnetic field. The calculation is done 
at finite temperature and the fermionic self energy is not supposed to 
be momentum-independent, which is the usual 
simplification in such calculations.
The phase diagram in the Temperature-Magnetic
field plane is determined. 
For each value of the magnetic field the critical temperature 
is higher than in the constant mass approximation. In
addition, the latter approximation shows a relative B-independence 
of the critical temperature, which occurs for stronger magnetic fields
in the momentum dependent case.

\newpage
 
\section{Introduction}

The mechanism of dynamical mass generation in the
presence of a magnetic field is particularly important
in connection with the vacuum structure of non-abelian field
theories such as QCD; also the QED case is
interesting both as a theoretical ``laboratory" and because of
possible applications.
In particular, scenaria of dynamical gauge symmetry breaking
in three-dimensional QED~\cite{farak} lead to interesting
and unconventional superconducting properties of the theory
after coupling to electromagnetism~\cite{mavromatos},
and therefore may be
of interest to the condensed matter physics, especially
in connection with the high-$T_c$ superconductors.

The phenomenon of magnetic catalysis has been studied by several
groups \cite{miransky,fkmm,klimenkoetsaklic} in 
various models. 
It has been found that a homogeneous magnetic field
induces  dynamical mass generation even for the weakest attractive
interaction between the fermions. In addition, in 2+1 dimensions the
mass generation is not restricted to a small number of flavours (which
is the case in the absence of the magnetic field). 

The treatment of the problem necessarily involves simplifications;
a serious one is the constant mass approximation, according to which
the fermionic self-energy is supposed to be momentum-independent. 
There has already been a first attempt \cite{afk} to treat the problem 
of dynamical mass generation
for QED in 2+1 and 3+1 dimensions at $T=0$, taking into account the 
momentum dependence of the fermion self-energy. The results have shown that
there are differences (very important ones in 3+1 dimensions) showing up when
the momentum dependence is taken into account, thus establishing the
necessity to go beyond the constant mass approximation. In that work,
attention was restricted to the strong magnetic field regime, to
facilitate the calculations. In the present work we will extend the
calculations in \cite{afk} to the finite temperature case in 2+1 dimensions.
Extensions to thermal 3+1 dimensional QED will be deferred to a
forthcoming publication. In the present work, only the regime of strong 
magnetic fields will be
studied, to render the problem tractable.

\section{Fermions in a constant magnetic field}

To fix our notations we shortly review here the characteristics of
fermions in a constant external magnetic field in 2+1
dimensions at zero temperature. 
The model we are going to consider is described by the Lagrangian density:

\be
{\cal L}=-\frac{1}{4}F_{\mu\nu}F^{\mu\nu}+i\ol\Psi D_\mu\gamma^\mu\Psi 
-m\ol\Psi\Psi,
\ee

\nin where $D_\mu = \partial_\mu+i g a_\mu + i e A^{ext}_\mu,$ 
$a_\mu$ is an abelian quantum gauge field, $F_{\mu \nu}$ is the
corresponding field strength, and $A^{ext}_\mu$ describes 
possible external fields; in this work $A^{ext}_\mu$ will 
represent a constant homogeneous external magnetic field. 
Notice that the fermions feel both the quantum and the external 
gauge fields, however we have allowed for different coupling 
constants, g and e, in order to give an effective description
of condensed matter systems \cite{mavromatos}. 
We recall the usual definition $g^2\equiv 4\pi\alpha$.

We will choose the ``symmetric" gauge for the external field

\be\label{symgauge}
A^{ext}_0(x)=0,~~A^{ext}_1(x)=-\frac{B}{2}x_2,
~~A^{ext}_2(x)=+\frac{B}{2}x_1
\ee

\nin for which we know from the work of Schwinger 
\cite{schwinger} that the fermion propagator is given in
the Minkowski space by the expression:

\be
S(x,y)=e^{iex^\mu A^{ext}_\mu(y)}\tilde S(x-y),
\label{sphase}
\ee

\nin where the translational invariant propagator $\tilde S$ 
has the following Fourier transform:

\bea\label{schwingerrep}
\tilde S(p)&=&\int_0^\infty ds e^{is\left(p_0^2-p_\bot^2
\frac{\tan(|eB|s)}{|eB|s}-m^2\right)}\nonu
&\times&
\left[(p^0\gamma^0+m)(1+\gamma^1\gamma^2\tan(|eB|s))-p^\bot\gamma^\bot
(1+\tan^2(|eB|s))\right],
\eea

\nin where $p^\bot=(p^1,p^2)$ is the transverse momentum and 
a similar notation holds for the $\gamma$ matrices.

Let us now turn to the finite temperature case.
We will denote the fermionic Matsubara frequencies by $\hat w_l=(2l+1)\pi T$ 
and the bosonic ones by $w_l=2l\pi T$. 
The translational invariant part
of the bare fermion propagator can be expressed in Euclidean space
by performing the rotations $p_0\to i\hat w_l$ and $s\to -is/|eB|$:

\bea\label{freeTprop}
\tilde S_l(p_\bot)&=&-\frac{i}{|eB|}
\int_0^\infty ds e^{-\frac{s}{|eB|}\left(\hat w_l^2+p_\bot^2
\frac{\tanh s}{s}+m^2\right)}\nonu
&\times&
\left[(-\hat w_l\gamma^3+m)(1-i\gamma^1\gamma^2\tanh s)
-p^\bot\gamma^\bot(1-\tanh^2s)\right]
\eea

\nin The Euclidean $\gamma$ matrices satisfy the 
anticommutation relations 
$\{\gamma^\mu,\gamma^\nu\}=-2\delta^{\mu\nu}$, with $\mu,\nu=1,2,3$.

The fermion propagator has another representation in the
lowest Landau level (LLL) approximation (\cite{chodos},\cite{miransky}),  
which at finite temperature reads:

\be
\tilde S_l(p_\bot)=-ie^{-p_\bot^2/|eB|}\frac{-\hat w_l\gamma^3+m}
{\hat w_l^2+m^2}
\left(1-i\gamma^1\gamma^2sign(eB)\right).
\label{sfree}
\ee

\nin Inspired by (\ref{sfree}) and taking into account the fact that
the dressed fermion propagator has the same phase dependence on $A^{ext}$ 
as the bare one \cite{afk}:
\be
G(x,y)=e^{iex^\mu A^{ext}_\mu(y)}\tilde G(x-y),
\label{gphase}
\ee
\nin we make the following ansatz for the Fourier transform of the
translational invariant part of the full propagator:

\be\label{lll}
\tilde G_l(p_\bot)=-ie^{-p_\bot^2/|eB|}\frac{-Z_l\hat w_l\gamma^3
+M_l}{Z_l^2\hat w_l^2+M_l^2}
\left(1-i\gamma^1\gamma^2sign(eB)\right)
\ee

\nin where $Z_l$ is the wave function renormalization and $M_l$ the
dynamical mass of the fermion. Both quantities depend only on the Matsubara 
index; this is a consequence of the restriction to the strong field
regime, where the LLL approximation reduces the fermion dynamics
to 1-dimensional dynamics \cite{mir3}. 
We will study the dynamical generation
of fermionic mass only in the case $m=0$.

\section{Integral equations and the recursion formula}

We can write the equations satisfied by the wave function renormalization
$Z_n$ and the dimensionless dynamical mass $\mu_n=M_n/\sqrt{|eB|}$ 
from the corresponding relations obtained from the Schwinger-Dyson 
equation at $T=0$ (\cite{shpagin}, \cite{afk}) making
the substitutions 
$$p_3\to \hat w_l~~~ {\rm and}~~~\int \frac{dp_3}{2\pi}\to T\sum_l$$ 
to obtain

\bea\label{intequaT}
Z_n&=&1-\frac{2\tilde\alpha t}{\hat\omega_n}
\sum_l\frac{Z_l\hat \omega_l}{Z_l^2\hat \omega_l^2+\mu_l^2}
\int_0^\infty rdr e^{-r^2/2}{\cal D}_{n-l}(r)\nonu
\mu_n&=&2\tilde\alpha t\sum_l\frac{\mu_l}{Z_l^2\hat \omega_l^2+\mu_l^2}
\int_0^\infty rdr e^{-r^2/2}{\cal D}_{n-l}(r)
\eea

\nin where we introduced the notations $\tilde\alpha=\alpha/\sqrt{|eB|}$,
$t=T/\sqrt{|eB|}$, $\hat\omega_l=\hat w_l/\sqrt{|eB|}$ and
$r$ stands for the dimensionless
modulus of the transverse momentum: $r^2=q_\bot^2/|eB|$. ${\cal D}_m(r)$ 
is the dimensionless longitudinal component ($\mu,\nu=3,3$) of the
photon propagator $D_{\mu\nu}$ which is the only one to play a r\^ole
since in the LLL approximation 
the fermion propagator is proportional to 
the projector operator $P=(1-i\gamma^1\gamma^2sg(eB))/2$ and the original 
Schwinger-Dyson equation contains the product \cite{afk}

\be
P\gamma^\mu P\gamma^\nu P D_{\mu\nu}=-P D_{33}
\ee

\begin{figure}
\epsfxsize=10cm
\epsfysize=8cm
\centerline{\epsfbox{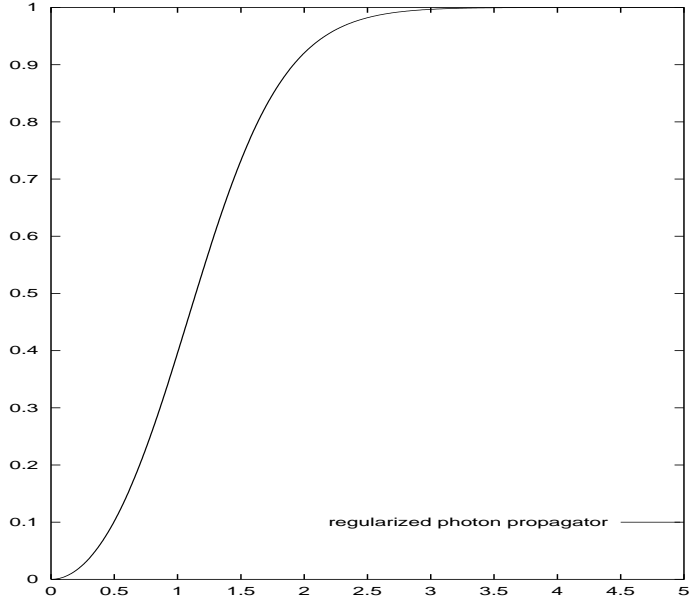}}
\caption{$(r^2+\omega_n^2) {\cal D}_n(r)$ 
versus $r$ for $n=0.$}
\label{propfig}
\end{figure}

\nin
We note that for $l=n$ in (\ref{intequaT}) the integration over 
$r$ is divergent in the infrared if we use the bare
photon propagator. Therefore we will use the   
{\em dressed} photon propagator, which is the subject 
of the next section. 
In figure \ref{propfig} we show the quantity 
$(r^2+\omega_n^2) {\cal D}_n(r)$ versus
$r.$ For the bare photon propagator this quantity is 1; in the figure we 
have set $n=0$ (for which the bare photon propagator leads to 
a divergence).
We observe that the dressed propagator equals the
bare one for most of the range of $r$ (from about 3 to infinity). 
The difference of the two is restricted to a small neighborhood
of $r=0.$ This is how this propagator cures the infrared 
divergencies associated with the use of the bare photon propagator. 

We will calculate the longitudinal 
component of the polarization tensor using the full Schwinger 
representation (\ref{schwingerrep}) of the fermion propagator since 
in the LLL approximation it is zero, both at $T=0$ and $T>0$.
This computation will be valid for any magnitude of the
magnetic field and will thus include the strong field approximation.
We only need the longitudinal component of the polarization 
tensor since we have \cite{kapusta} (with the dimensional 
reduction from 3+1 to 2+1)

\be
{\cal D}_n(r)=
\frac{1}{r^2+\omega_n^2-F}P_L^{33}+
\frac{\omega_n^2}{(r^2+\omega_n^2)^2}
\ee

\nin with 

\be
P_L^{33}=\frac{r^2}{r^2+\omega_n^2}~~~~\mbox{and}~~~~
F=\frac{\tilde\Pi_n^{33}(r)}{P_L^{33}}~~~~\mbox{where}~~~~
\tilde\Pi_n^{33}(r)=\frac{\Pi_n^{33}(\sqrt{|eB|}r)}{|eB|},
\ee

\nin which leads to

\be \label{Dprop}
{\cal D}_n(r)
=\frac{1}{r^2+\omega_n^2}+
\frac{r^2\tilde\Pi_n^{33}(r)}{(r^2+\omega_n^2)^2
\left(r^2-\tilde\Pi_n^{33}(r)\right)}
\ee

\nin in the {\it Feynman gauge} that will be used in this paper.

To solve (\ref{intequaT}), we will proceed as in the case $T=0$ \cite{afk}:
we put a trial series $\mu_n^{trial}$ in the integral 
equations and obtain a first 
estimate $\mu_n^{(1)}$ of the series $\mu_n$. 
We then use this series in the integral equation and build up a 
(converging) iterative procedure. The trial series will be a solution of
the recursion formula that we derive now, following a procedure similar
to the one that gave the differential equation at zero 
temperature in \cite{miransky}, \cite{afk}.
We start by splitting the summation over $l$ and write, starting from 
equation (\ref{intequaT}) satisfied by $\mu_n:$

\bea
\mu_n&\simeq& 2\tilde\alpha t\sum_{|l|\le n}
\frac{\mu_l}{Z_l^2\hat\omega_l^2+\mu_l^2}
\int_0^\infty rdr e^{-r^2/2}{\cal D}_n(r)\nonu
&+& 2\tilde\alpha t\sum_{|l|>n}\frac{\mu_l}{Z_l^2\hat\omega_l^2+\mu_l^2}
\int_0^\infty rdr e^{-r^2/2}{\cal D}_l(r)
\eea

\nin such that the difference $\mu_{n+1}-\mu_n$ reads 

\be
\mu_{n+1}-\mu_n=2\tilde\alpha tf_n
\sum_{|l|\le n}\frac{\mu_l}{Z_l^2\hat\omega_l^2+\mu_l^2}
\label{fe}
\ee

\nin where we defined

\be
f_n=\int_0^\infty rdr e^{-r^2/2} \left[{\cal D}_{n+1}(r)-
{\cal D}_n(r)\right]
\ee

\nin Since we wish to get rid of the sum in equation (\ref{fe}),
we eliminate it manipulating the expressions for the differences 
$\mu_{n+1}-\mu_n$ and $\mu_{n+2}-\mu_{n+1}.$
We finally obtain:

\be\label{recursion}
\frac{\mu_{n+2}}{f_{n+1}}-\left(\frac{1}{f_{n+1}}+\frac{1}{f_n}\right)
\mu_{n+1}+\frac{\mu_n}{f_n}=
4\tilde\alpha t\frac{\mu_{n+1}}{Z_{n+1}^2\hat\omega_{n+1}^2+\mu_{n+1}^2}
\ee 

\begin{figure}
\epsfxsize=10cm
\epsfysize=8cm
\centerline{\epsfbox{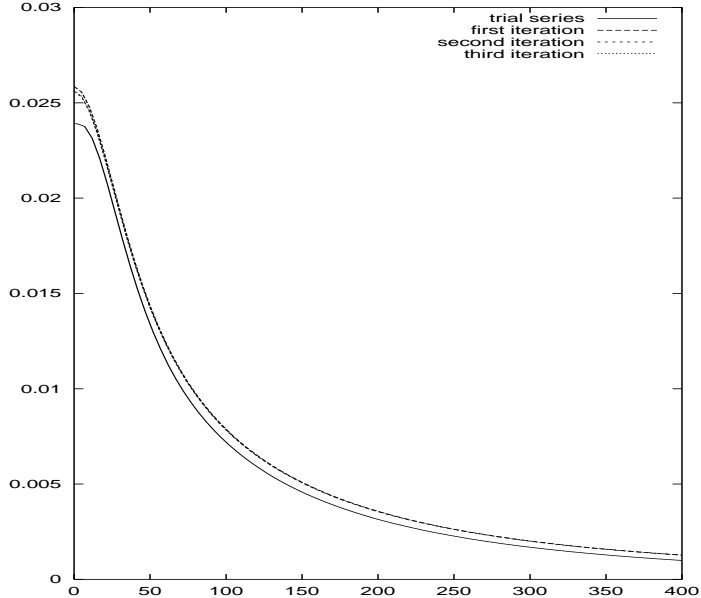}}
\caption{$\mu_n$ versus $n$ for
$\tilde\alpha=.01$ and $t=.001$}
\label{iteram}
\end{figure}

\begin{figure}
\epsfxsize=10cm
\epsfysize=8cm
\centerline{\epsfbox{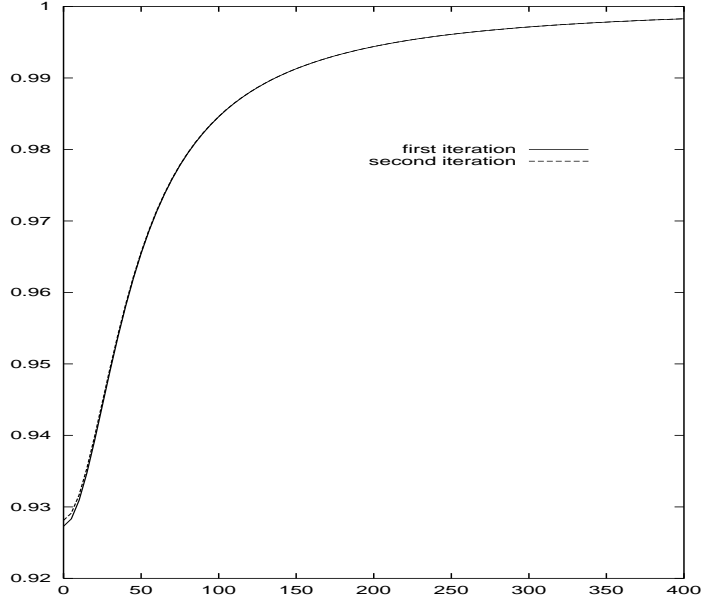}}
\caption{$Z_n$ versus $n$ for
$\tilde\alpha=.01$ and $t=.001$}
\label{iteraz}
\end{figure}

\nin The solution of (\ref{recursion}) is found by giving (equal) initial
values to $\mu_0$ and $\mu_1,$ such that $\lim_{n\to\infty}\mu_n=0$. 
We plot in figure \ref{iteram} the solution of (\ref{recursion})
(where we take $Z_n=1$ for every $n$)
as well as the first iterations in the integral equations (\ref{intequaT}).
We see that the convegence of the iterative procedure is very quick.
We show in figure \ref{iteraz} the wave function renormalization
versus $n$, for which the convergence is also very quick. 
From the technical point of view, the function $f_n$ has been computed
with a Gauss-Hermite quadrature of order 40 and the series $\mu_n$ and
$Z_n$ have been truncated to an order 
depending on the temperature,  
adjusted in such a way that a given precision 
was kept throughout the numerical computations.

\section{Longitudinal polarization tensor}

We now proceed with the calculation of the polarization tensor 
using the Schwinger 
representation of the fermion propagator (\ref{freeTprop}).

The one-loop polarization tensor is 

\be\label{poltensor}
\Pi^{\mu\nu}_n(k_\bot)=
-4\pi\alpha T\sum_l\int\frac{d^2p_\bot}{(2\pi)^2}
\mbox{tr}\left\{\gamma^\mu \tilde S_l(p_\bot)
\gamma^\nu\tilde S_{l-n}(p_\bot-k_\bot)\right\}
\ee

\nin
We note that the $A^{ext}_\mu$-dependent phase of the fermion propagator  
does not contribute to the polarization tensor since in coordinate space
this phase contribution is

\be
\exp\left\{ie\left(x^\mu A^{ext}_\mu(y)+
y^\mu A^{ext}_\mu(x)\right)\right\}=1
\ee

\nin as can be seen from the potential (\ref{symgauge}). 
We also remark that, 
since we use the full expression for the fermion propagator, 
the result of this calculation will be valid 
{\em for any value} of the 
external magnetic field. 
 
Using the expression (\ref{freeTprop}) of the fermion propagator, we obtain
for the longitudinal component,
after the integration over the transverse momentum $p_\bot$

\bea
\Pi^{33}_n(k_\bot)&=&\frac{-4\alpha T}{|eB|}\sum_l\int_0^\infty
\frac{dsd\sigma}{\tanh s+\tanh\sigma}
e^{-\frac{k_\bot^2}{|eB|}\frac{\tanh s\tanh\sigma}{\tanh s+\tanh\sigma}
-\frac{1}{|eB|}[(s+\sigma)(\hat w_l^2+m^2)+
\sigma w_n(w_n-2\hat w_l)]}\nonu
&&\times
\left[k_\bot^2\frac{\tanh s\tanh\sigma}{(\tanh s+\tanh\sigma)^2}-
\left(\frac{|eB|}{\tanh s+\tanh\sigma}\right)
(1-\tanh s)(1-\tanh\sigma)\right.\nonu
&&+\left(\hat w_l(\hat w_l-w_n)-m^2\right)
\left(1+\tanh s\tanh\sigma\right)\Bigg]
\eea

\nin where we will take the fermionic dynamical mass for $m$ and
consider that it does not depend on the momentum. 
We use this approximation since it is the only way to keep 
the calculation of the polarization tensor tractable. Without this 
assumption the momentum integrals cannot be done analytically;
the numerical treatment on the other hand would be prohibitively
demanding. 

As in \cite{tsai}, we make the change of variable  $s=u(1-v)/2$ and
$\sigma=u(1+v)/2$ to obtain

\bea\label{pol1}
\Pi^{33}_n(k_\bot)&=&
\frac{-2\alpha T}{|eB|}\sum_l\int_0^\infty udu\int_{-1}^1 dv
e^{-\frac{k_\bot^2}{|eB|}\frac{\cosh u-\cosh uv}{2\sinh u}-
\frac{u}{|eB|}[\tilde\omega_l^2+m^2
+(1+v)w_n(w_n/2-\tilde\omega_l)]}\nonu
&\times&\left[\left(\tilde\omega_l(\tilde\omega_l-w_n)
-m^2\right)\coth u
-\frac{|eB|}{\sinh^2 u}+k_\bot^2\frac{\cosh u-\cosh uv}{2\sinh^3 u}
\right]
\eea

\nin We integrate by parts:

\be
\int_0^\infty du e^{-\phi(u)}\frac{u}{\sinh^2 u}\longrightarrow
\int_0^\infty du
e^{-\phi(u)}\coth u\left(1-u\frac{d\phi(u)}{du}\right), 
\ee

\nin where we have discarded the surface term \cite{tsai}.
This term is finite 
but would lead to an infinite summation over the Matsubara modes.
We then obtain the final expression:

\bea\label{pol2}
&&\Pi^{33}_n(k_\bot)=\frac{-2\alpha}{|eB|}\int_0^\infty du\int_{-1}^1 dv
e^{-\frac{k_\bot^2}{|eB|}\frac{\cosh u-\cosh uv}{2\sinh u}-\frac{u}{|eB|}
[m^2+\frac{1-v^2}{4}w_n^2]}\\
&\times&T\sum_le^{-\frac{u}{|eB|}W_l^2}
\left[u\frac{k_\bot^2}{2}\frac{\cosh uv-v\coth u\sinh uv}{\sinh u}
+\coth u\left(2uW_l^2+uvw_nW_l-|eB|\right)\right]\nonumber
\eea

\nin where $W_l=\hat w_l-\frac{(1+v)}{2}w_n$.

At this point we must make an important remark. In equation (\ref{pol2})  
there is a potential divergence of the integral over $u$ coming from
the sum involving $|eB|\coth u.$ We find out that if we perform 
the summation over the Matsubara modes {\it before} performing the integration,
this would-be divergence cancels against the sum involving
$2u W_l^2 \coth u.$ This can be easily seen using the Poisson resummation
\cite{dittrich}:

\be\label{poiss}
\sum_{l=-\infty}^\infty e^{-a(l-z)^2}=
\left(\frac{\pi}{a}\right)^{1/2}
\sum_{l=-\infty}^\infty e^{-\frac{\pi^2l^2}{a}-2i\pi zl},
\ee

\nin which shows that the difference below has no divergence since

\be\label{ressum}
T\sum_{l=-\infty}^\infty\left(2uW_l^2-|eB|\right)
e^{-\frac{uW_l^2}{|eB|}}
=\frac{|eB|^{5/2}}{2\sqrt\pi u^{3/2}T^2} 
\sum_{l\ge 1} (-1)^{l+1}l^2 e^{-\frac{|eB|l^2}{4uT^2}}
\cos[\pi nl(1+v)]
\ee

\nin so that the integration over $u$ in (\ref{pol2}) is safe, both on the 
infrared and the ultraviolet sides. The conclusion 
is that one should {\it first} 
sum over the Matsubara modes and perform the integral 
over $u$ {\it afterwards.}

If we take the limit $T \to 0$ in (\ref{pol2}), 
we recover the zero-temperature results given 
in \cite{fkmm} since the substitutions
$W_l\to p_3$ and $T\sum_l\to (2\pi)^{-1}\int dp_3$ lead to 

\bea \label{tfirst} 
\lim_{T\to 0}&&T\sum_l e^{-\frac{u}{|eB|}W_l^2}=
\frac{1}{2}\sqrt\frac{|eB|}{\pi u}\nonu
\lim_{T\to 0}&&T\sum_l \left(2uW_l^2-|eB|\right)
e^{-\frac{u}{|eB|}W_l^2}=0\nonu
\lim_{T\to 0}&&T\sum_l uvw_nW_l e^{-\frac{u}{|eB|}W_l^2}=0
\eea

\nin such that $(\omega_n\to k_3)$

\bea\label{piindep}
&&\lim_{T\to 0}\Pi_n^{33}(k_\bot)=
\frac{-\alpha}{2\sqrt{\pi|eB|}}
\int_0^\infty du\sqrt u\int_{-1}^1 dv
e^{-\frac{k_\bot^2}{|eB|}\frac{\cosh u-\cosh uv}{2\sinh u}-\frac{u}{|eB|}
[m^2+\frac{1-v^2}{4}k_3^2]}\nonu
&&~~~~~~~~~~~~~~~~~~~~~~~~~~~~~~\times k_\bot^2
\frac{\cosh uv-v\coth u\sinh uv}{\sinh u}
\eea

\nin
We call the attention of the reader to a rather tricky 
aspect of these limiting procedures. 
In equations (\ref{tfirst}) we have taken the limit $T \to 0$ {\em before}
we integrate over $u$  and the result has been consistent with the outcome
of the zero temperature calculation. However, one might equally well 
start from the expression (\ref{pol2}), and take the limit $T \to 0$
{\em after} the integration over $u$ has been performed. 
One may wonder whether the two limits are the same, that is whether the
operations of taking $T \to 0$ and integrating over $u$ commute.
It is easy to see that {\em they do not, 
unless we keep a non-zero fermion mass}. Let us see this considering 
the expression (\ref{pol2}) for $n=0$ and $k_\bot=0$:

\bea\label{nonzerom}
\Pi^{33}_0(0)&=&-2\alpha\int_0^\infty du\int_{-1}^1 dv
e^{-\frac{u}{|eB|}m^2}\\
&\times&T\coth u\sum_l e^{-\frac{u}{|eB|}\hat w_l^2}
\left(2\frac{u}{|eB|}\hat w_l^2-1\right)\nonumber
\eea

\nin We observe that, if $m$ is zero, the proper time $u$ 
in the integrand appears 
only in the combination $uT^2/|eB|$ (with the exception of 
$\coth u$). This suggests that we perform the natural change of variable 
$u\to u|eB|/T^2.$ The above expression becomes:

\bea\label{zerom}
\Pi^{33}_{0,m=0}(0)&=&-\frac{4\alpha|eB|}{T}\int_0^\infty du 
\coth\left(\frac{u|eB|}{T^2}\right)\nonu
&\times&\sum_le^{-u\pi^2(2l+1)^2}
\left(2u\pi^2(2l+1)^2-1\right)
\eea

\nin and shows that $\Pi^{33}_{0,m=0}(0)$ would 
be proportional to $1/T$ in the limit $T\to 0$ where 
we have $\coth(u|eB|/T^2)\simeq 1$. 
Thus we can interchange the limit $T\to 0$ and the integration 
over the proper time $u$ to find the correct zero temperature limit
only if $m \ne 0$, at least as long as $|eB|\ne 0$.
This condition is consistent since the magnetic field
always generates a dynamical mass when $T<T_c$. 

Let us take the zero magnetic field 
limit of (\ref{pol2}), after making the change of variable
$u\to |eB|u$. We obtain then:

\bea\label{limebzero}
\lim_{|eB|\to 0}\Pi^{33}_n(k_\bot)&=&
-2\alpha \int_0^\infty du\int_{-1}^1 dv
e^{-u[m^2+\frac{1-v^2}{4}(k_\bot^2+w_n^2)]}\nonu
&\times&T\sum_le^{-uW_l^2}
\left[\frac{k_\bot^2}{2}(1-v^2)+
\frac{1}{u}\left(2uW_l^2+uvw_nW_l-1\right)\right]
\eea

\nin The integration in (\ref{limebzero})
over the proper time $u$ gives the relevant result of  
\cite{mavromatos} (equation (A12),
after performing the momentum integration and changing the 
Feynman parameter $x$ into $(1-v)/2$). For the case at hand ($|eB|=0$), 
we can take a massless fermion ($m=0$) and the value of the photon 
thermal mass 
is found by setting first $n=0$ and then take the limit 
$k_\bot\to 0$ in (\ref{limebzero}):

\be\label{thermm}
M_{|eB|=0}^2(T)=-\lim_{k_\bot\to 0}\Pi^{33}_0(k_\bot)=c~\alpha T
\ee

\nin with

\be
c=4\int_0^\infty du
\sum_le^{-u(2l+1)^2}\left[2(2l+1)^2-\frac{1}{u}\right]
\ee 

\nin To compute $c$, we use the Poisson resummation (\ref{poiss})
and write

\bea
c&=&2\pi^{5/2}\int_0^\infty \frac{du}{u^{5/2}}
\sum_{l\ge 1}(-1)^{l+1}l^2e^{-\frac{\pi^2l^2}{4u}}\nonu
&=&\frac{16}{\sqrt\pi}\sum_{l\ge 1}\frac{(-1)^{l+1}}{l}
\int_0^\infty dx\sqrt xe^{-x} \nonu
&=&8\ln 2
\eea 

\nin which gives the result that was found in \cite{mavromatos} 
and \cite{aitchison} (with
the notation $\alpha\to\alpha/4\pi$).

Finally we decompose $\Pi_n^{33}(k_\bot)$ in a sum of two terms: 
the temperature independent part and the temperature dependent one.
Using equation (\ref{pol2}) and the Poisson resummation (\ref{poiss}), 
a straightforward computation leads to 

\be\label{pitot}
\Pi_n^{33}(k_\bot)=\Pi_n^0(k_\bot)+\Pi_n^T(k_\bot)
\ee

\nin where $\Pi_n^0(k_\bot)$ is the zero temperature part 
(\ref{piindep})
and $\Pi^T_n(k_\bot)$ the temperature dependent part:

\bea\label{pidep}
&&\Pi^T_n(k_\bot)=\frac{-\alpha}{\sqrt{\pi |eB|}}
\int_0^\infty du\int_{-1}^1 dv
e^{-\frac{k_\bot^2}{|eB|}\frac{\cosh u-\cosh uv}{2\sinh u}-\frac{u}{|eB|}
[m^2+\frac{1-v^2}{4}w_n^2]}\\
&&\times
\left[k_\bot^2\frac{\sqrt{u}}{\sinh u}(\cosh uv-v\coth u\sinh uv)
\sum_{l\ge 1}(-1)^le^{-\frac{|eB|l^2}{4uT^2}}\cos[\pi nl(1+v)]+\right.\nonu
&&\left.
\frac{\coth u}{\sqrt{u}}|eB|\sum_{l\ge 1}(-1)^{l+1}e^{-\frac{|eB|l^2}{4uT^2}}
\left(\frac{|eB|l^2}{uT^2}\cos[\pi nl(1+v)]+2nl\pi v\sin[\pi nl(1+v)]
\right)\right].
\nonumber
\eea

\nin We compute in appendix A the strong field $(|eB| \to \infty)$
asymptotic form (\ref{asypol}) of $\Pi^0+\Pi^T$ that we used for 
the numerical analysis of (\ref{intequaT}).

We note that the other components of the polarization tensor were 
computed \cite{jalex} in 3+1 dimensions and their computation in 
2+1 dimensions would follow the same steps.

Finally, we give the thermal photon mass, using the dimensionless
variables already introduced: 

\be\label{thermphotmass}
\mu_{phot}^2=-\lim_{k_\bot\to 0}
\frac{\Pi^{33}_0(k_\bot)}{|eB|}=\frac{2\tilde\alpha}{\sqrt\pi t^2}
\int_0^\infty du e^{-u\mu^2}\frac{\coth u}{u^{3/2}}\sum_{l\ge 1}
(-1)^{l+1}l^2e^{-\frac{l^2}{4ut^2}}
\ee

\begin{figure}
\epsfxsize=10cm
\epsfysize=8cm
\centerline{\epsfbox{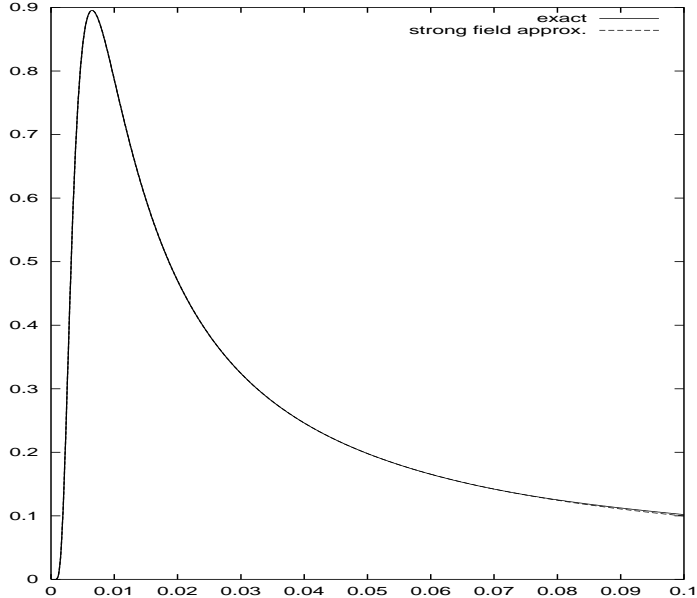}}
\caption{$\mu_{phot}^2$ versus $t$ for
$\tilde\alpha=.01$ and $\mu=.01$}
\label{thma}
\end{figure}

\nin where $\mu=m/\sqrt{|eB|}$. 
We are actually interested in the behaviour of $\Pi_0^{33}$
for all the values of the ratio $\mu/t$: $\mu/t\to 0$
for the description of the phase transition where the
dynamical mass vanishes for $t=t_c>0$ and $\mu/t\to\infty$
for the zero temperature limit and $\mu>0$.
In figure \ref{thma} we plot
$\mu_{phot}^2$ as a function of $t$ 
as well as its strong field asymptotic form obtained from the
equation (\ref{asypol}) of appendix A.

\bea\label{thmaapp}
\mu_{phot}^2&\simeq&4\frac{\tilde\alpha}{t}
\sum_{l\ge 1}(-1)^{l+1}le^{-l\mu/t}\nonu
&=&4\tilde\alpha\frac{d}{d\mu}\left(\sum_{l\ge 1}
(-e^{-\mu/t})^l\right)\nonu
&=&4\tilde\alpha\frac{d}{d\mu}
\left(\frac{-e^{-\mu/t}}{1+e^{-\mu/t}}\right)\nonu
&=&4\frac{\tilde\alpha}{t}\frac{e^{-\mu/t}}{(1+e^{-\mu/t})^2}
\eea

\nin 
We can see the perfect
agreement between the curves for the whole range of the ratio
$\mu/t$, as long as $\mu<<1$ and $t<<1$ which is the case we study 
in the framework of the strong field approximation. 
We note the unusual behaviour of the thermal photon mass:
for decreasing temperature  
$\mu_{phot}^2$ {\it increases} as $1/t$ 
as long as $t>\mu$, reaches a maximum 
and then decreses to 0 when $t<<\mu$. The result is a 
large correction to the photon propagator when $t\simeq\mu$.

\section{Phase diagram}

We wish to study the critical temperature and the critical field
of the theory,
defined by the relations $m_{dyn}(B,T_c(B))=0$ and $m_{dyn}(B_c(T),T)=0$.
 
\begin{figure}
\epsfxsize=10cm
\epsfysize=8cm
\centerline{\epsfbox{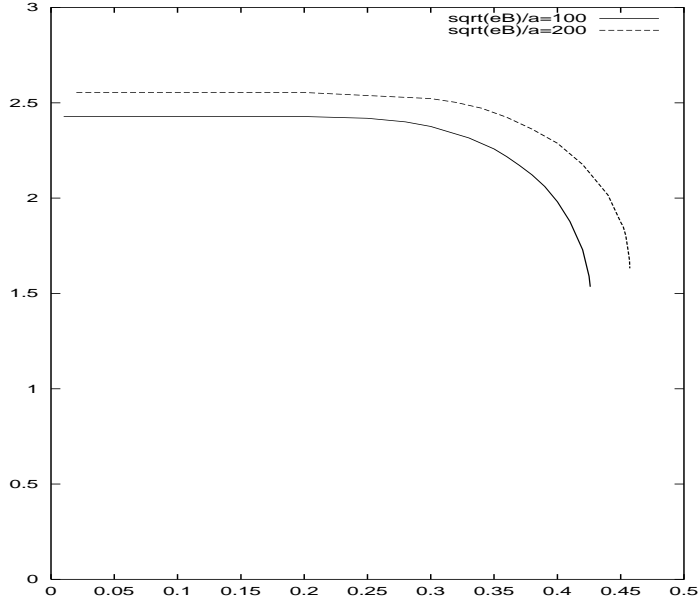}}
\caption{$m_{dyn}/\alpha$ versus $T/\alpha$}
\label{dmt}
\end{figure}

Let us first find the critical temperature when we fix the magnetic
field. We plot in figure
\ref{dmt} the evolution of the dimensionless dynamical mass 
$m_{dyn}/\alpha=\mu_{n=0}/\tilde\alpha$ with the dimensionless
temperature $T/\alpha$. We start from $T=0$ and
see that $m_{dyn}$ first
follows a plateau for
small temperatures and then decreases when $T$ reaches
a value that we interpret as the critical temperature,  
$T_c.$ We cannot find any other solution than $m_{dyn}=0$ when $T>T_c$
and the vertical slope 
when $T=T_c$ suggests 
that the transition might be of first order. 
We note that the recursion formula (\ref{recursion}) gives a
second solution which is not physical since it does not lead to a
converging iterative procedure in the integral equation (\ref{intequaT})
and that this unphysical solution also does not go beyond $T=T_c$.
Finally, we noticed that 
the convergence of the iterative procedure to solve the Schwinger-Dyson
equation is slower as we approach the critical temperature.
In the same figure we have plotted the results for two magnetic fields; the
resulting dynamical mass and critical temperature 
are bigger for the biggest magnetic field, as 
expected.

\begin{figure}
\epsfxsize=10cm
\epsfysize=8cm
\centerline{\epsfbox{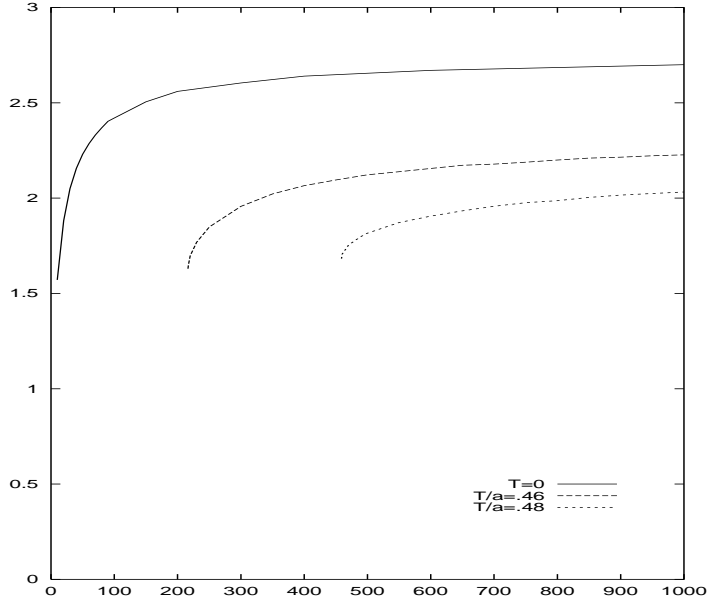}}
\caption{$m_{dyn}/\alpha$ versus $|eB|^{1/2}/\alpha$} 
\label{dmb}
\end{figure}

Let us now study the dynamical mass when we fix the temperature.
We plot in figure \ref{dmb} the evolution of 
$m_{dyn}/\alpha$ as a function of 
$\sqrt{|eB|}/\alpha$ fixing the ratio $T/\alpha$.
We see that for a given non-zero temperature, there is a minimal value 
that the field must take to generate a dynamical mass, that is, 
a critical field.
In the case $T=0$, we see that no critical field is needed to generate
a dynamical mass, as is known \cite{miransky} (we didn't plot 
the curve down to $|eB|=0$ since the 
strong magnetic field approximation is not valid in this region).
Notice that the calculation of this curve, which has been done using 
the methods of this work (and taking the limit $T \to 0$) is identical 
to the corresponding result of the genuine $T=0$ calculation of \cite{afk}.
For $T>0$, the transition again seems to be of 
the first order and the convergence 
of the iterative procedure is slower as we approach the critical field.
If the temperature increases, the curve shows smaller dynamical mass and
higher critical magnetic field. 

\begin{figure}
\epsfxsize=10cm
\epsfysize=8cm
\centerline{\epsfbox{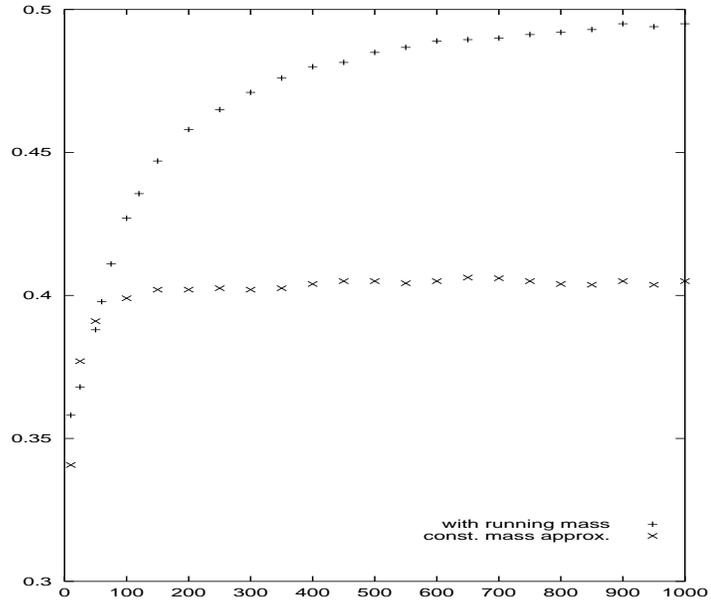}}
\caption{$T_c/\alpha$ versus $|eB|^{1/2}/\alpha$}
\label{phase}
\end{figure}

We depict in figure \ref{phase} the phase diagram
$T_c/\alpha$ versus $\sqrt{|eB_c|}/\alpha$
and make the comparison with the constant mass approximation 
(where $\mu_l$ does not depend on $l$) discussed in appendix B.
We see that the constant mass approximation systematically 
underestimates the critical temperature $T_c(B)$
and that this one is almost independent 
from the magnetic field for $|eB|^{1/2}/\alpha > 100.$  
Going beyond the constant mass approximation 
we find that the critical temperature increases with $|eB|$
and reaches asymptotically the value $T_c^\infty=\alpha/2$ 
when $|eB|\to\infty$. 
The fact that the temperature increases with increasing magnetic
field is consistent with similar studies made in the Gross-Neveu 
model \cite{klimenko}.

\begin{figure}
\epsfxsize=10cm
\epsfysize=8cm
\centerline{\epsfbox{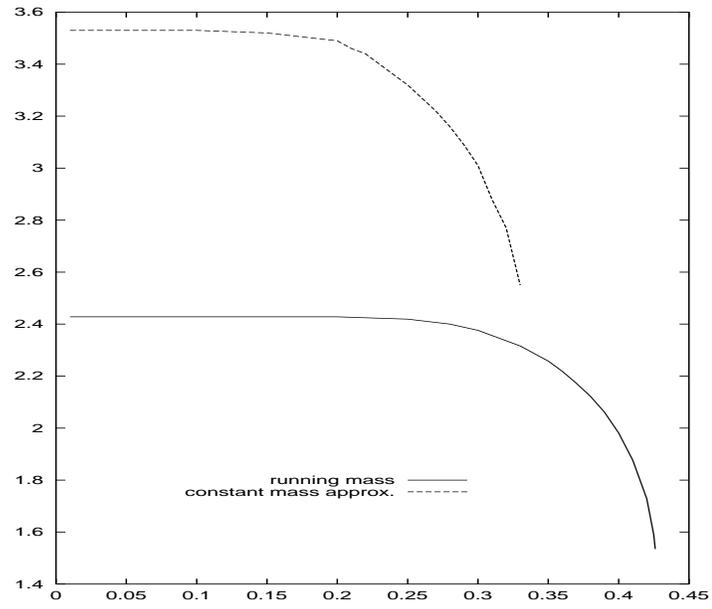}}
\caption{$m_{dyn}/\alpha$ versus $T/\alpha$ for $|eB|^{1/2}/\alpha=100$}
\label{dmtconst}
\end{figure}

It is interesting to compare the momentum-dependent case with the 
constant mass approximation along paths of constant temperature 
or constant magnetic field.
In figure \ref{dmtconst} we compare the constant
mass approximation and the momentum dependent dynamical mass for a given 
value of the magnetic field. We see in the constant mass approximation
that the dynamical mass is overestimated, 
as was found in \cite{afk}, but that 
the critical temperature is smaller than in the momentum dependent study.

\begin{figure}
\epsfxsize=10cm
\epsfysize=8cm
\centerline{\epsfbox{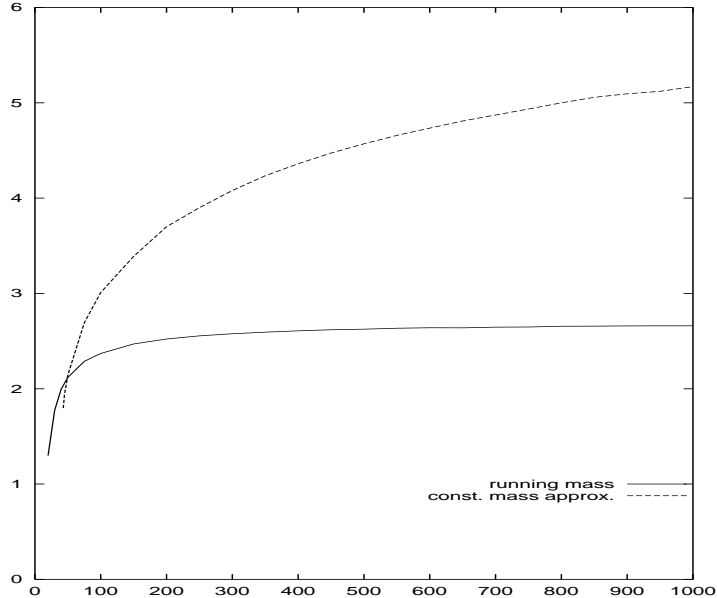}}
\caption{$m_{dyn}/\alpha$ versus $|eB|^{1/2}/\alpha$ for  
$T/\alpha=0.3.$} 
\label{dmbconst.fig}
\end{figure}

In figure \ref{dmbconst.fig} we fix the temperature 
and plot the dynamical mass $m_{dyn}/\alpha$ as a function of 
the magnetic field for both the momentum-dependent case and the constant 
mass approximation. The constant mass approximation is seen to 
overestimate the dynamical mass for most of the range of the magnetic field;
in addition the critical magnetic field turns out to be bigger for this
approximation than for the momentum-dependent case.
An important remark should be made here. 
For sufficiently high magnetic field strength the critical temperature is
almost independent from the magnetic field. This means that 
if the temperature is set to a sufficiently large value,
it will not be possible to find a corresponding critical field. 
Incidentally, this was the reason why we have chosen a 
safely low value for the temperature in figure \ref{dmbconst.fig}.
The results of figures \ref{dmtconst} and \ref{dmbconst.fig},
are consistent with the phase diagram, where it was found that
the constant mass approximation
underestimates the critical temperature (for constant magnetic field) and
overestimates the critical magnetic field (for constant temperature).

\begin{figure}
\epsfxsize=10cm
\epsfysize=8cm
\centerline{\epsfbox{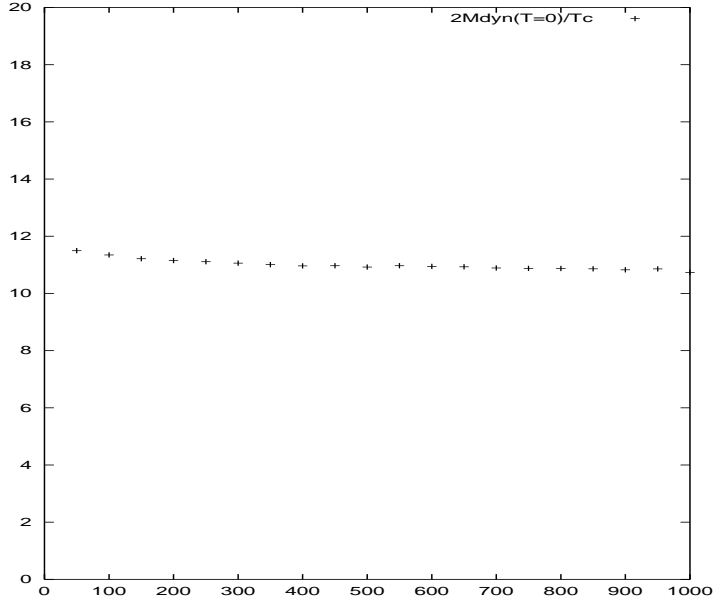}}
\caption{Ratio $2m_{dyn}(T=0)/T_c$ versus $|eB|^{1/2}/\alpha$}
\label{comp}
\end{figure}

Finally we show in figure \ref{comp} the ratio $2m_{dyn}(T=0)/T_c$ 
versus $\sqrt{|eB|}/\alpha$, where $m_{dyn}(T=0)$ is the
dynamical mass that is obtained at the limit $T=0$.
This ratio is an important parameter for superconductivity theories.
We find out that this ratio is of order 10 and 
approximately constant over the whole range of
$\sqrt{|eB|}/\alpha$
considered here; this is consistent with the result 
obtained before in \cite{mavromatos} where $|eB|=0$,
which shows that the magnetic field has no important influence
on this ratio. 

As a side remark, we computed the condensate 

\be
\left<0|\overline\psi\psi|0\right>=\lim_{x\to y}
\mbox{tr}G(x,y)=\mbox{tr}T\sum_l\int
\frac{d^2p_\bot}{(2\pi)^2}\tilde G_l(p_\bot)
\ee

\nin which is another order parameter for this transition.
The LLL approximation gives:

\bea
\left<0|\overline\psi\psi|0\right>&=&
-4i\int\frac{d^2p_\bot}{(2\pi)^2}e^{-p_\bot^2/|eB|}
\times iT\sum_l \frac{M_l}{Z_l^2\hat w_l^2+M_l^2}\nonu
&=&\frac{|eB|}{\pi}T\sum_l \frac{M_l}{Z_l^2\hat w_l^2+M_l^2}
\eea

\nin and we noticed numerically that 
$<0|\overline\psi\psi|0>/|eB|$ almost 
does not depend on the temperature or the magnetic field, taking 
into account the momentum dependence of the self energy. When $T>T_c$
or $|eB|<|eB_c|$,
the condensate is zero since the only solution for the series $M_l$ 
is $M_l=0$ for all $l$.
The constant mass approximation is obtained
very easily with the method of summation described in appendix B
and the result reads

\be\label{constcond}
\left<0|\overline\psi\psi|0\right>=
\frac{|eB|}{2\pi}\tanh\left(\frac{\tilde\beta\mu_0}{2}\right)
\ee

\nin which shows that the condensate vanishes at the critical temperature
or the critical field, when $\mu_0=0$.
We note that an analytical computation of the condensate
beyond the LLL approximation and the constant mass approximation 
has been performed in \cite{fkm}, leading to 
(\ref{constcond}) when $|eB|\to\infty$.

\section{Discussion}

The effect of external magnetic fields on the state of the condensate
is of great interest, in view of recent experiments
with high-$T_c$ cuprates pertaining to the
thermal conductivity of quasiparticle excitations
in the superconducting state~\cite{ong}. It is argued that 
quasiparticle thermal conductivity
plateaux in a high-magnetic-field phase of these materials
indicate the opening of a gap for strong magnetic fields,
depending on the intensity of the external field, when the latter
is applied perpendicularly to the cuprate planes.
The phenomenon appears to be generic
for systems of charged Dirac fermions in external magnetic fields,
in the sense that strong enough magnetic fields are capable of
inducing spontaneous formation of neutral condensates,
whose magnitude scales with the magnetic field
strength~\cite{miransky,hong,sc,ng}.
Such gaps disappear {\em at critical temperatures
of the order of the gap.} 

The authors of \cite{ong} find that, for strong enough magnetic fields
of ${\cal O}[1]-{\cal O}[10]$ Tesla, there are plateaux in the
thermal conductivity of quasiparticle excitations
about the $d$-wave state (in particular about the nodes
of the d-wave superconducting gap). Such plateaux are  interpeted
as an indication of the opening of a new gap, 
induced by the magnetic fields
at the nodes. These plateaux disappear at a critical
temperature that depends on the magnetic field intensity, and in particular
they report the empirical relation:
\be
T^{exp}_c \propto \sqrt{|eB|}
\label{empirical}
\ee
for the dependence of the observed critical temperature
on the external magnetic field strength.

The mass gap (that is, the
dynamical mass at zero temperature) has been found in \cite{afk} to
be linear in $\sqrt{|eB|},$ so the critical temperature is expected to
be linear in $\sqrt{|eB|}.$
The above has been a conjecture from the zero temperature case, but
a genuine finite temperature calculation had to be done before one 
could draw definite conclusions. 
The relevant results are contained in figure \ref{phase}
and we may conclude the following: 

\begin{itemize}

\item In the constant mass approximation the
critical temperature $T_c/\alpha$ is constant (its value is about 0.4) 
for a wide range of the magnetic field. The critical temperature depends
on the magnetic field only for relatively small field strengths. We
note that this calculation has been done assuming a strong magnetic field, 
so one cannot explore the region of very small fields in a reliable way.
Note that because of the choice of axes
in figure \ref{phase} a square root dependence would show up as a straight
line, which may be identified with the tangent to the curve. 
Thus the possibility that the behaviour of the critical temperature 
on the magnetic field conforms to the experimental result (\ref{empirical}) 
is restricted to a small range of magnetic field strengths (approximately 
up to $\sqrt{|e B|}/\alpha \simeq 50).$  

\item In the case of the momentum-dependent fermionic self energy 
the situation changes quantitatively: a straight line, corresponding to a 
square-root-like dependence fits to the data up to the value
$\sqrt{|eB|}/\alpha \simeq 150.$ 
Then the curve levels off and $T_c$ approaches $0.5 \alpha.$ 
which is somewhat bigger than the strong field limit in the 
constant mass approximation. However it is not clear what this 
``saturation" phenomenon implies for the superconductor case: if 
the field strength becomes very big, it will move the system out of 
the superconducting phase, so the whole calculation is not relevant any more.
It appears that one should stay at a safe distance from both the 
``too weak" field region (where the LLL approximation collapses) and the ``too
strong" field regime, which the physical phenomenon may 
disappear {\footnote {We thank the referee for pointing this out.}}. 
For not too extreme fields the square root empirical relationship appears to 
persist in the momentum dependent case for relatively stronger fields 
than it is possible for the constant mass approximation.

\end{itemize}

\section*{Acknowledgements}

This work has been done within the
TMR project ``Finite temperature phase transitions in particle Physics",
EU contract number: FMRX-CT97-0122.
The authors would like to acknowledge financial support from the
TMR project.

\begin{appendix}
\section{Strong field approximation for the 
longitudinal polarization tensor}

Starting from the relations (\ref{piindep}) and (\ref{pidep}), 
we compute here the asymptotic 
forms of $\Pi^0$ and $\Pi^T$ in the strong field limit where the dominant 
contribution in the integration over $u$ comes from the region $u>>1$,
which can be seen with the change of variable $u\to u|eB|$.
We will use the dimensionless parameters previously introduced. 
For the temperature independent part, we can take $\mu=0$ and 
we write that for $u>>1$:

\bea
\frac{\cosh uv-v\coth u\sinh uv}{\sinh u}&\simeq&
(1-v)e^{-u(1-v)}+(1+v)e^{-u(1+v)}\nonu
e^{-\frac{k_\bot^2}{|eB|}\frac{\cosh u-\cosh uv}{2\sinh u}}&\simeq&
e^{-\frac{k_\bot^2}{2|eB|}}
\eea

\nin such that we obtain from (\ref{piindep})

\bea\label{asypolindep}
\Pi_n^0(k_\bot)\simeq&&\frac{-\tilde\alpha}{2\sqrt\pi}
k_\bot^2 e^{-\frac{k_\bot^2}{2|eB|}}
\int_0^\infty du\int_{-1}^1dv\sqrt ue^{-u\frac{1-v^2}{4}\omega_n^2}
\left[(1-v)e^{-u(1-v)}+(1+v)e^{-u(1+v)}\right]\nonu
=&&\frac{-\tilde\alpha}{\sqrt\pi}
k_\bot^2 e^{-\frac{k_\bot^2}{2|eB|}}
\int_{-1}^1dv\frac{1-v}{\left(1-v+\frac{1-v^2}{4}\omega_n^2\right)^{3/2}}
\times\int_0^\infty du\sqrt ue^{-u}\nonu
=&&-2\sqrt 2\tilde\alpha k_\bot^2 e^{-\frac{k_\bot^2}{2|eB|}}
\frac{1}{2+\omega_n^2}
\eea

\nin For the temperature dependent part, we have to keep $\mu\ne 0$
for the reason previously explained and we will consider the 
dominant term in the strong field approximation, i.e.

\bea
\frac{\Pi^T_n(k_\bot)}{|eB|}&\simeq&\frac{-\tilde\alpha}{\sqrt\pi t^2}
\sum_{l\ge 1}(-1)^{l+1}l^2\int_0^\infty du \int_{-1}^1 dv
e^{-\frac{k_\bot^2}{|eB|}\frac{\cosh u-\cosh uv}{2\sinh u}-
u(\mu^2+\frac{1-v^2}{4}\omega_n^2)}\\
&&~~~~~~~~~~~~\times
\frac{\coth u}{u^{3/2}}e^{-\frac{l^2}{4ut^2}}\cos[\pi nl(1+v)]\nonu
&\simeq&\frac{-\tilde\alpha}{\sqrt\pi t^2}e^{-\frac{k_\bot^2}{2|eB|}}
\sum_{l\ge 1}(-1)^{l+1}l^2\int_{-1}^1 dv\cos[\pi nl(1+v)]
\int_0^\infty \frac{du}{u^{3/2}}e^{-u(\mu^2+\frac{1-v^2}{4}\omega_n^2)
-\frac{l^2}{4ut^2}}\nonumber
\eea

\nin We first have the following integration over $u$ which leads
to the Bessel function $K_{1/2}$:

\be
I(a,b)=\int_0^\infty \frac{du}{u^{3/2}}e^{-au-\frac{b}{u}}=
2\left(\frac{a}{b}\right)^{1/4}K_{1/2}(2\sqrt{ab})=
\sqrt\frac{\pi}{b}e^{-2\sqrt{ab}}
\ee

\nin with $a=\mu^2+\frac{1-v^2}{4}\omega_n^2$ and $b=\frac{l^2}{4t^2}$. 
Then we approximate the integration over $v$ by interpolating 
between the asymptotic expression of the integral when $n>>1$ 
and the one with $n=0$. When $n>>1$, the oscillations of the 
integrand make the contributions of opposite sign
cancel, such that the dominant contribution
for the integral comes from the region  
$1-\frac{1}{2|n|l}\le v\le 1$ where
the cosine does not vanish:

\bea
J_n(\mu,t)&=&\int_{-1}^1 dv\cos[\pi nl(1+v)]
e^{-\frac{l}{t}\sqrt{\mu^2+\frac{1-v^2}{4}\omega_n^2}} \nonu
&\simeq& 2\int_{1-\frac{1}{2|n|l}}^1dv \cos[\pi nl(1+v)]e^{-\frac{l\mu}{t}}
=\frac{2}{\pi |n|l}e^{-\frac{l\mu}{t}}  
\eea

\nin We also have $J_0(\mu,t)=2e^{-\frac{l\mu}{t}}$ and therefore we
make the ansatz: 

\be
J_n(\mu,t)\simeq\frac{2e^{-\frac{l\mu}{t}}}{1+\pi |n|l}
\ee

\nin With these approximations, we can finally write

\be\label{asypoldep}
\frac{\Pi^T_n(k_\bot)}{|eB|}\simeq \frac{-4\tilde\alpha}{t}
e^{-\frac{k_\bot^2}{2|eB|}}\sum_{l\ge 1}(-1)^{l+1}
\frac{l}{1+\pi |n|l}e^{-\frac{l\mu}{t}}
\ee
 
\nin The strong field asymptotic longitudinal polarization tensor is then,
from (\ref{asypolindep}) and (\ref{asypoldep})

\be\label{asypol}
\frac{\Pi_n^{33}(k_\bot)}{|eB|}\simeq -2\tilde\alpha
e^{-\frac{k_\bot^2}{2|eB|}}\left[\sqrt 2\frac{k_\bot^2}{|eB|}
\frac{1}{2+\omega_n^2}+\frac{2}{t}\sum_{l\ge 1}(-1)^{l+1}
\frac{l}{1+\pi |n|l}e^{-\frac{l\mu}{t}}\right]
\ee

\section{Constant mass approximation}

We compute here explicitly the summation over the Matsubara modes in 
the integral equation (\ref{intequaT}) for the dynamical mass in 
the approximation where the latter does not depend on the Matsubara
index $l:~\mu_l=\mu_0$. 
We suppose that the correction to the photon propagator also
does not depend on the Matsubara index and keep its transverse
momentum dependence only.
The integral equation (\ref{intequaT}) for the dynamical mass 
with the help of (\ref{Dprop}) yields (we take $Z_l=1$): 

\be
1=2\tilde\alpha\int_0^\infty r dr e^{-r^2/2}
\left[\Sigma_t^1(r,\mu_0)-
\frac{r^2\mu_t^2(r)}{r^2+\mu_t^2(r)}\Sigma_t^2(r,\mu_0)\right]
\ee

\nin where $\mu_t^2(r)=\frac{-1}{|eB|}\Pi_0^{33}(r^2|eB|)$
and the sums over Matsubara modes are

\bea
\Sigma_t^1(r,\mu_0)&=&t\sum_l\frac{1}{(\hat\omega_l^2+\mu_0^2)
[(\hat\omega_l-\hat\omega_0)^2+r^2]}\nonu 
\Sigma_t^2(r,\mu_0)&=&t\sum_l\frac{1}{(\hat\omega_l^2+\mu_0^2)
[(\hat\omega_l-\hat\omega_0)^2+r^2]^2}
\eea

\nin $\Sigma_t^1$ has been computed in \cite{gusyninT} and we 
shortly repeat here the steps of the computation.
$\Sigma_t^1$ can be expressed
in terms of the contour integral

\be
\Sigma_t^1(r,\mu_0)=
\frac{1}{2i\pi}\int_C d\omega\frac{1}{1+e^{-\tilde\beta\omega}}
\frac{1}{(\mu_0^2-\omega^2)[r^2-(\omega-i\hat\omega_0)^2]}
\ee

\nin The contour $C$ runs around the poles $i(2l+1)\pi t$ of the
Fermi-Dirac distribution function with residues equal to $1/t$.
This contour can be deformed to run around the 4 poles $\pm \mu_0$
and $\pm r+i\hat\omega_0$ of the rational fraction. One must
not forget that this deformation leads to an
overall minus sign since we travel along the new contour
in the clockwise direction. The summation
over the 4 corresponding residues gives then

\bea\label{matsusum1}
\Sigma_t^1(r,\mu_0)&=&
\frac{1}{2\mu_0}\tanh\left(\frac{\mu_0\tilde\beta}{2}\right)
\frac{\hat\omega_0^2+r^2-\mu_0^2}{(\hat\omega_0^2+r^2-\mu_0^2)^2+
4\mu_0^2\hat\omega_0^2}\nonu
&+&\frac{1}{2r}\coth\left(\frac{r\tilde\beta}{2}\right)
\frac{\hat\omega_0^2-r^2+\mu_0^2}{(\hat\omega_0^2-r^2+\mu_0^2)^2+
4r^2\hat\omega_0^2}                                                     
\eea

\nin where $\tilde\beta=\beta\sqrt{|eB|}=1/t$. 

$\Sigma_t^2$ is simply given by
$\Sigma_t^2(r,\mu_0)=
-\frac{1}{2r}\frac{\partial}{\partial r}\Sigma_t^1(r,\mu_0),$
so:

\bea\label{matsusum2}
\Sigma_t^2(r,\mu_0)&=&
\frac{1}{2\mu_0}\tanh\left(\frac{\tilde\beta\mu_0}{2}\right)
\frac{(r^2-\mu_0^2+\hat\omega_0^2)^2-4\mu_0^2\hat\omega_0^2}
{[(r^2-\mu_0^2+\hat\omega_0^2)^2+4\mu_0^2\hat\omega_0^2]^2}\\
&&+\frac{\tilde\beta}{8r^2\sinh^2(\frac{\tilde\beta r}{2})}
\frac{\hat\omega_0^2-r^2+\mu_0^2}
{(\hat\omega_0^2-r^2+\mu_0^2)^2+4r^2\hat\omega_0^2}\nonu
&&+\frac{1}{4r^3}\coth\left(\frac{r\tilde\beta}{2}\right)
\frac{\hat\omega_0^2-r^2+\mu_0^2}
{(\hat\omega_0^2-r^2+\mu_0^2)^2+4r^2\hat\omega_0^2}\nonu
&&+\frac{1}{2r}\coth\left(\frac{r\tilde\beta}{2}\right)
\frac{4\hat\omega_0^2(\hat\omega_0^2+\mu_0^2)-
(\hat\omega_0^2-r^2+\mu_0^2)^2}
{[(\hat\omega_0^2-r^2+\mu_0^2)^2+4r^2\hat\omega_0^2]^2}\nonumber
\eea

\nin 
The critical temperature $T_c$ in the constant mass approximation satisfies 
the equation

\be\label{constmass}
1=2\tilde\alpha\int_0^\infty r dr
e^{-r^2/2}\left[\Sigma_{t_c}^1(r,0)-
\frac{r^2\mu_{t_c}^2(r)}{r^2+\mu_{t_c}^2(r)}\Sigma_{t_c}^2(r,0)\right]
\ee

\nin with

\bea
\Sigma_{t_c}^1(r,0)&=&\frac{\tilde\beta_c}{4}\frac{1}{r^2+\hat\omega_c^2}
+\frac{1}{2r}\coth\left(\frac{r\tilde\beta_c}{2}\right)
\frac{\hat\omega_c^2-r^2}{(\hat\omega_c^2+r^2)^2}\nonu
\Sigma_{t_c}^2(r,0)&=&\frac{\tilde\beta_c}{4}
\frac{1}{(r^2+\hat\omega_c^2)^2}
+\frac{\tilde\beta_c}{8r^2\sinh^2(\frac{\tilde\beta_c r}{2})}
\frac{\hat\omega_c^2-r^2}{(\hat\omega_c^2+r^2)^2}\nonu
&&+\frac{1}{4r^3}\coth\left(\frac{r\tilde\beta_c}{2}\right)
\frac{\hat\omega_c^2-r^2}{(\hat\omega_c^2+r^2)^2}
+\frac{1}{2r}\coth\left(\frac{r\tilde\beta_c}{2}\right)
\frac{3\hat\omega_c^2-r^2}{(\hat\omega_c^2+r^2)^3}\nonumber
\eea

\nin where $\hat\omega_c=\pi t_c$ and $\mu_{t_c}^2(r)$ is the 
opposite of the strong field asymptotic form of the dimensionless 
longitudinal polarization tensor (\ref{asypol})
for zero Matsubara index taken at the critical temperature:

\bea
\mu_{t_c}^2(r)&=&
\tilde\alpha e^{-r^2/2}\left[\sqrt 2r^2+\frac{4}{t_c}
\lim_{\mu\to 0}\sum_{l\ge 1}(-1)^{l+1}l e^{-l\mu/t}\right]\nonu
&=&\tilde\alpha e^{-r^2/2}\left[\sqrt 2r^2+\frac{1}{t_c}\right]
\eea

\nin where the summation over $l$ has been done to
obtain equation (\ref{thmaapp}). 
The solution of (\ref{constmass}) is plotted in figure \ref{phase}.

\end{appendix}

\end{document}